\documentstyle[twocolumn,epsf,aps,prl]{revtex}
\begin{document}
\title{White-Scalapino-like Stripes in a Mean-Field Hubbard Model}
\author{C. Kusko and R.S. Markiewicz}
\address{Physics Department and Barnett Institute, 
Northeastern U.,
Boston MA 02115}
\maketitle
\begin{abstract}
A mean field calculation of the Hubbard model finds a rich phase diagram. 
The antiferromagnetic phase is generally unstable away from half filling, and 
there are several regions of phase separation.  One solution in particular 
closely resembles the stripe phase of White and Scalapino.  By comparison to 
unrestricted Hartree-Fock calculations (for which this phase is metastable), 
it is demonstrated that this phase arises from phase separation.  
The interface surface tension is found to change sign below a particular stripe 
width, at which point the stripes begin to meander, gradually crossing over to 
diagonal.
\end{abstract}

\narrowtext

When the Hubbard model is doped away from the antiferromagnetic insulator at
half filling, a number of calculations find evidence for spatially 
inhomogeneous solutions.  There is considerable debate\cite{Su,CCGB} as to 
whether these solutions are generic features of the Hubbard model, or arise only
in a restricted parameter domain.  Related issues are whether the inhomogeneity 
is driven by phase separation or antiferromagnetic (AFM) domain wall 
formation\cite{Cod}, and how these features are related to `stripes' in 
cuprates and other oxides.  Similarly, in the doped tJ model, various 
calculations find that the ground state is striped\cite{WhiSc}, or 
uniform\cite{PC}, or phase separated\cite{HelMan,pry}.  Inclusion of realistic 
values of $t^{\prime}$ into the model further reduces stripe 
stability\cite{Tohy}.  Reference to earlier calculations may be found in these 
articles and in the reviews\cite{PhSRev}.  

Unrestricted Hartree-Fock (UHF) calculations\cite{HF,VLLGB,KMNF} find that the
holes form filled (one additional hole per row) stripes which act as antiphase
boundaries between AFM domains.  Such filled domain wall stripes are not found
in more advanced calculations\cite{WhiSc,HelMan} of the Hubbard model, and are
not consistent with experiment on the cuprates\cite{Tran}.  We here
analyze a {\it metastable} state of the UHF calculations, which closely
resembles the White-Scalapino (WS)\cite{WhiSc} stripes, and agrees better with
experiment.  These stripes can be understood from a phase separation approach, 
comparing the free energies of low-order commensurate magnetic phases, $q_x,q_y
\sim$ 0 or $Q_i=\pi /a$.  The resulting mean-field phase diagrams involve phase 
separation between the AFM
phase and a metallic phase, either ferromagnetic (FM), as in early ferron phase 
approaches to the Hubbard model\cite{Nag}, or a phase resembling  
WS stripes, depending on the value of second 
neighbor hopping parameter $t'$.  These stripes are stable local free energy 
minima in UHF calculations, but globally there are alternative states of lower 
free energy\cite{KMNF}.  However, these solutions can be stabilized by 
{\it additional interactions} beyond the pure Hubbard model (e.g., 
charge-density wave or superconducting), and hence may be relevant to
experiment.  These additional interactions will be discussed in a 
companion publication\cite{MK}; here we introduce the mean-field model
and utilize UHF calculations to calculate the surface tension in the resulting 
stripe phases.  We find that WS-like stripes are stable against macroscopic 
phase separation.

We study a one-band electron-hole symmetric Hubbard model [interaction = $U\sum_
i(n_{i\uparrow}-1/2)(n_{i\downarrow}-1/2)$] with bare dispersion
$\epsilon_k=-2t(c_x+c_y)-4t'c_xc_y$,
with $c_i=\cos{k_ia}$.  In the presence of a mean-field magnetization $m_q$ at 
wave vector $\vec q$, the quasiparticle dispersion becomes
\begin{equation}
E_{\pm}={1\over 2}(\epsilon_k+\epsilon_{k+q}\pm E_0),
\label{eq:3}
\end{equation}
where $E_0=\sqrt{(\epsilon_k-\epsilon_{k+q})^2+4U^2m_q^2}$.
The site magnetization is found self-consistently from
\begin{equation}
m_q=\sum_k (f(E_-)-f(E_+)){Um_q\over E_0},
\label{eq:6}
\end{equation}
with Fermi function $f(E)=1/(1+e^{(E-E_F)/k_BT})$.  The free energy is
\begin{equation}
F=\sum_{k,i=\pm}E_if(E_i)-TS+U(m_q^2+{x^2\over 4}),
\label{eq:7}
\end{equation}
where $S$ is the entropy.

\begin{figure}
\leavevmode
   \epsfxsize=0.41\textwidth\epsfbox{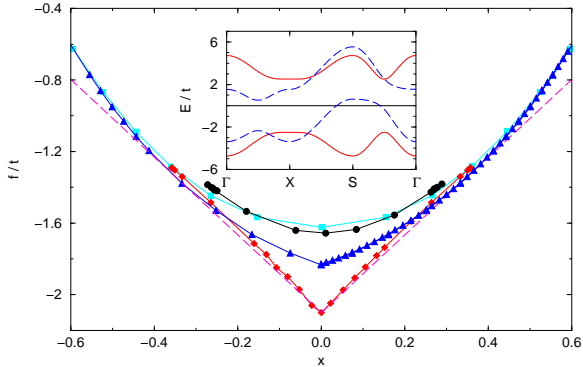}
\vskip0.5cm 
\caption{Free energy vs. doping for several magnetic phases of the Hubbard model
($U=6.03t$, $t'=0$).  Diamonds = AFM, triangles = LAF, circles = FM, and 
squares = PM phase.  Dashed lines = tangent construction. Inset:
Dispersion of magnetic phases: solid lines = AFM at $x=0$, dashed lines = 
LAF at $x=0.353$; Brillouin zone points 
$\Gamma =(0,0)$, $X=(\pi ,0)$, $S=(\pi ,\pi$).}
\label{fig:1}
\end{figure}

\begin{figure}
\leavevmode
   \epsfxsize=0.35\textwidth\epsfbox{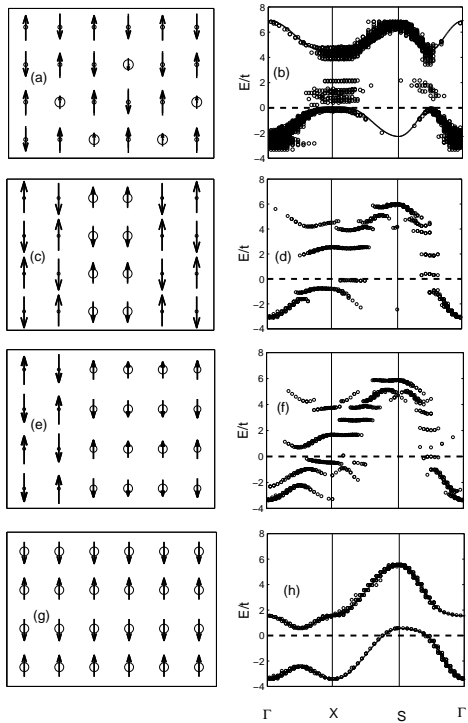}
\vskip0.5cm 
\caption{Unrestricted Hartree-Fock calculations of the Hubbard
model. Left panels: the hole configurations for 1/8, 1/6, 2/9, and 1/3
dopings minimizing the free energy. Right panels: the corresponding
dispersions (open circles); solid lines = mean-field bands for undoped 
antiferromagnet (b) and LAF (h); dashed lines = chemical potential. 
(a,b): Random distribution of holes in an
antiferromagnetic background forming magnetic polarons at 1/8 doping. 
(c,d): LAF stripes at 1/6 doping  - on the charge stipes the hole
density is 0.27 and the magnetization is 0.46; on the
antiferromagnetic stripes the hole density is 0.06 and the
magnetization is 0.77. (e,f): AFM stripes in LAF background at x=2/9; there is a
weak modulation of the hole density on the LAF stripe: 0.245 (0.365) holes on 
the outer (inner) rows.  (g,h): LAF configuration at 1/3 doping.}
\label{fig:12}
\end{figure}

The phase with $\vec q=\vec Q\equiv (\pi ,\pi )$ provides a good model for the 
AFM phase at half filling with Mott gap, 
successfully describing the spin wave dispersion\cite{SWZ} and Monte Carlo 
results\cite{BSW,GEH}, and serving as the basis for a number of treatments of 
strong correlation effects\cite{ChM}.  A fit to the dispersion of the magnetic 
insulator SrCuO$_2$Cl$_2$ finds\cite{OSP} $t=325meV$, $U=6.03t$, and $t'=-0.276
t$, which will be assumed below unless stated otherwise.  We use the same model,
with different choices of $\vec q\ne\vec Q$, to describe a number of competing 
magnetically ordered states.  While at half filling the AFM state has lowest 
free energy, this is not true for finite doping $x$, leading to a rich phase 
diagram, with regimes of phase separation.  
Figure~\ref{fig:1} shows the low-temperature free energy as a function of doping
for the case $t'=0$ for three magnetic phases, the standard antiferromagnet 
(AFM) $\vec q=\vec Q$, a ferromagnet (FM) with $\vec q=(0,0)$, and a linear 
antiferromagnet (LAF) with $\vec q=(\pi ,0)$ (see Fig.~\ref{fig:12}g, below).
The curves are symmetric about half filling ($x=0$).  For $|x|\le 0.25$ the AFM 
lies lowest in energy; between $0.25\le |x|\le 0.65$ the LAF lies lowest, and 
beyond that, the ground state is nonmagnetic ($m_q=0$).  For all dopings, the FM
state is metastable.  At high doping the magnetic phases terminate when 
$m_q\rightarrow 0$. The inset to Fig.~\ref{fig:1} shows the dispersions for the 
stable phases, AFM at $x=0$ (solid lines) and LAF at $x=0.353$ (dashed lines).  

The antiferromagnetic state has a cusp at half filling, with the slope 
discontinuity being the Mott-Hubbard gap.  Away from half-filling, this state is
always {\it thermodynamically unstable} -- the compressibility $\sim\partial^2f
/\partial x^2$ is negative.  The tangent construction (dashed lines) shows that 
the equilibrium state between zero doping and $|x|=x_c=0.353$ consists of a {\it
phase separation} between the AFM and LAF phases.  Note that the mean-field
model misses the true UHF ground state, which has filled ($x=1$) stripes in an 
AFM background.  It can be shown that, if the last term in Eq.~\ref{eq:7} is 
omitted, this ground state would be recovered for large $U$, with the LAF phase 
stable only for a small parameter range near $U/t=6-8$ ($t'=0$).  

\begin{figure}
\leavevmode
   \epsfxsize=0.40\textwidth\epsfbox{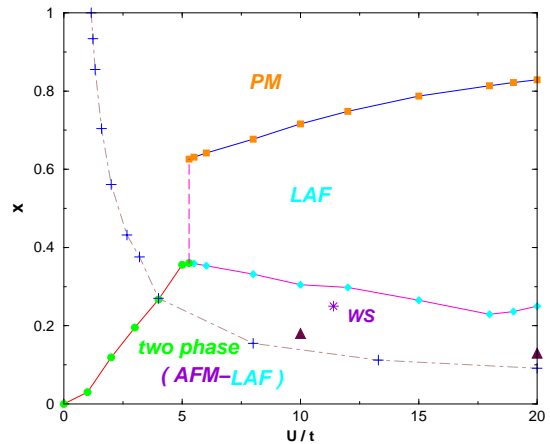}
\vskip0.5cm 
\caption{Phase diagram, $x(U)$ for the Hubbard model, with $t'=0$. 
Triangles = Hubbard model results, estimated from Fig. 1
of Ref.~\protect\onlinecite{CCGB}; dot-dashed line with $+$'s = tJ model 
results, Ref.~\protect\onlinecite{HelMan}b, assuming $J/t=4t/U$.}
\label{fig:6}
\end{figure}

The resulting phase diagram $x$ vs $U$ is shown in Fig.~\ref{fig:6}.
Phase separation persists for all finite $U$, but while the insulating state is
always AFM, there is a crossover in the metallic stripe component from 
paramagnetic phase for $U<U_c=5.3t$ to LAF for $U>U_c$).  
When $t'\ne 0$, the phase diagram is completely different, with phase separation
between the AFM and a FM phase\cite{MK}.
For large $U$, the Hubbard 
model should reduce to the tJ model; agreement with recent calculations for the 
phase separation boundary in the Hubbard\cite{CCGB} and tJ\cite{HelMan} models 
is satisfactory (triangles and $+$'s in Fig.~\ref{fig:6}a).  The deviation at 
small $U$ (large $J$) is expected, since the models are equivalent only in the 
large-$U$ limit.  While the metallic phase in the tJ model is usually taken as 
paramagnetic, the WS results may hint that it is an LAF phase near $U=11t$.

Figure~\ref{fig:12} illustrates some of the low-energy textures found 
in UHF calculations, and shows that in the LAF stripe phase dispersions 
the added states form additional bands near midgap, as found in ordered stripe 
arrays\cite{OSP} and for randomly distributed magnetic polarons 
(Fig.~\ref{fig:12}a,b). 
[For the left-hand panels of Fig.~\ref{fig:12}, the UHF calculations were 
iterated to self-consistency on 24$\times$24 (a), 32$\times$6 (c), or 
12$\times$12 (e,g) lattices with periodic boundary conditions.  For the 
dispersions of the right-hand panels of Fig.~\ref{fig:12}, these solutions were 
extended to a 32$\times$32 (d,h) or 36$\times$36 (f) lattice, with one 
additional iteration (Fig.~\ref{fig:12}b was on a 24$\times$24 lattice).]  

The LAF stripes resemble the WS stripes of the tJ model: the 
minimum LAF stripe is two cells wide, and acts as an antiphase boundary between 
AFM domains, Fig.~\ref{fig:12}c.  In both calculations, the doped ground state 
is found to involve mixtures of LAF and AFM stripes, with no sign of insulating,
empty stripes.  The doping is comparable: the star in Fig.~\ref{fig:6}a 
represents the WS stripe, assuming an effective $U/t=4t/J$, with $J=0.35t$.  
Both kinds of stripe have similar fractional transfer of holes onto adjacent 
AFM rows (see caption of Fig.~\ref{fig:12}), and both are destabilized by 
non-zero $t'$.  We find that the charged stripes have a fixed, minimal width for
$x\le 1/6$, with the charge per row of a stripe doubling at higher doping, and 
the stripe phase terminating near $x=1/3$; WS find similar doping dependences,
systematically shifted due to the difference in hole density (1/3 vs 1/4) on a 
stripe.  Similar LAF stripes were 
found previously as metastable UHF solutions\cite{VLLGB}.  An LAF-like state has
also been found in recent Monte Carlo calculations in the manganites\cite{AY};  
interestingly, a spin flux phase can form from a coherent superposition of two 
LAF phases. 

\begin{figure}
\leavevmode
   \epsfxsize=0.40\textwidth\epsfbox{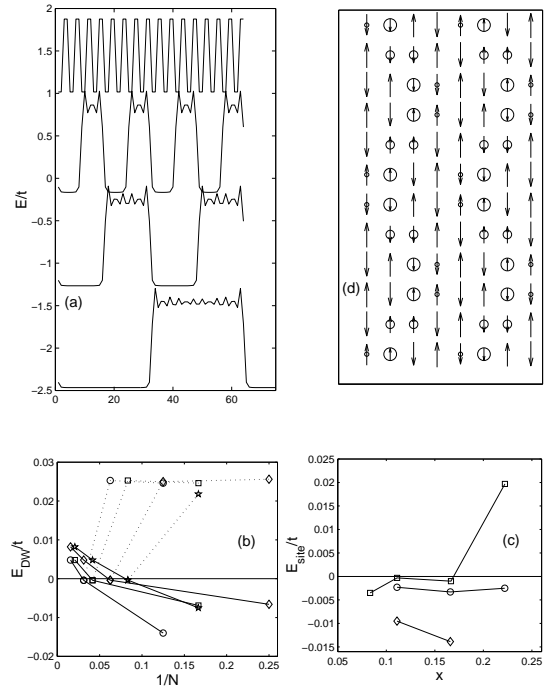}
\vskip0.5cm 
\caption{(a) Free energy $f(x)$ per row for a LAF stripe 
array in the Hubbard model, with $t'=0$, $U=8t$ and $x$=1/6, comparing several 
different lattice periodicities: 32 LAF $\times$ 32 AFM for a 128$\times$6 
lattice, and, for 64$\times$6 lattices, 16$\times$16, 8$\times$8, and 2$\times
$2.  Energies of narrower arrays are offset for clarity. (b) Excess free energy 
{\it per domain wall atom} $E_{DW}$ plotted vs $1/N$, where $N$ is the charge 
stripe periodicity, for $x$ = 1/12 (circles), 1/9 (squares), 1/6 (diamonds), and
2/9 (stars). Dotted lines connect unstable vertical stripes.  (c) Comparison of 
energy per site $E_{site}$ of meandering (circles) and 
diagonal (squares) stripes compared to the filled diagonal stripes 
(diamonds)\protect\onlinecite{KMNF}. 
(d) Pattern of spin and charge order on meandering $x$=1/6 stripe.}
\label{fig:11}
\end{figure}

It is clear from Fig.~\ref{fig:12} that the LAF-AFM stripes arise from phase
separation, with the stripe spacing evolving as expected with doping.  However
the question remains as to whether the stripes are stable against macroscopic
phase separation -- i.e., is the free energy of the stripe phase $F_{str}$ 
higher or lower than that of the separated bulk phases (tie-line) $F_{sep}$?  
We answer this via UHF calculations, taking care to minimize finite size 
effects.  This is done by (a) adjusting $U$ so that $x_0$ is a simple rational 
fraction (at $U/t=8$, $x_0\simeq 1/3$), and (b) working with large lattices, up 
to $128\times 6$.  Figure~\ref{fig:11} shows (a) the average free energy on each
row of a series of AFM-LAF stripe arrays, of the same average doping ($x=1/6$) 
but different stripe widths and (b) the resulting surface tension $\sigma$ 
(free energy difference $F_{str}-F_{sep}$ per domain wall atom).  
The magnetic contribution, associated with excess holes pushed onto the magnetic
bounding layers, is positive and saturates for wider stripes, while the LAF 
energy is negative, and oscillates on the stripes, in parallel with hole density
oscillations.  These oscillations are due to quantum confinement, similar to the
Friedel oscillations seen in electrons confined on a step on a Cu 
surface\cite{Crom}.  The confinement oscillations lead to a long-range 
interaction between domain walls (across a charged stripe), which explains why 
the net surface tension saturates so slowly as a function of stripe width, and 
why it depends mainly on the width of the LAF stripes.

For the widest stripes, the surface tension $\sigma$ starts to level off to a 
value of $\sim 0.1t$ per domain wall atom for an isolated domain wall.  
As the stripes move closer $\sigma$ decreases, ultimately changing sign 
(negative surface tension).  When the LAF stripe has a width of 8 cells, the
surface tension is essentially zero.  For narrower LAF stripes straight vertical
stripes are unstable, but can be pinned by commensurability effects on specially
chosen lattices; the free energy is generally high (dotted lines in 
Fig.~\ref{fig:11}b).  On larger lattices, the UHF spontaneously evolves to a
meandering stripe pattern, Fig.~\ref{fig:11}d, with free energy lower than the
tieline (Fig.~\ref{fig:11}b, lowest points of solid lines).  The meandering LAF 
stripes are composed of straight diagonal segments, separated by kinks.  On the 
straight segments, holes on successive rows are shifted diagonally by one Cu 
site, leading to ferromagnetic alignment between kinks.  Remarkably, the
free energy of the meandering stripes is lower than that of the corresponding 
straight diagonal stripes, Fig.~\ref{fig:11}c (although the difference is very
small, and could be a finite size effect).  The crossover appears to be kinetic
energy driven: the holes in the LAF phase are delocalized along the (FM) rows,
but when the LAF stripes get too narrow, adjacent rows shift to provide a FM
coupling.  
Due to commensurability pinning effects, it will be hard to repeat this 
calculation for arbitrary values of $U$, although $x_0\sim$1/4 at $U=16t$.

There is a gradual crossover (Fig.~\ref{fig:11}c) from vertical (at $x$=2/9) to 
meandering (1/6, 1/9) to diagonal stripes (1/12).  At $x$=1/12, the diagonal 
stripes have a low free energy, and meandering configurations are unstable.  
Figure~\ref{fig:11}c also includes the free
energy of the diagonal, one hole per row stripes which are the UHF ground
state\cite{KMNF}.  The free energy differences are small, and the order of
states may be reversed by including some additional (perhaps phononic or
Coulomb) interactions.

Some mention must be made about the size of the lattice used.  Most of the
results correspond to lattices $96\times 6$ (for $x$ = 1/9 and 2/9), $128\times
6$ (for all $x$ = 1/12, and for the largest period at $x$ = 1/6), or $64\times 
6$ (for the remaining $x$ = 1/6), with periodic boundary conditions assumed.  
The meandering stripes were all on $48\times 12$ lattices, and the diagonal on
$24\times 24$ ($48\times 16$ for 1/12).  Straight stripes are metastable when 
the LAF stripe width is 4, and we had to use special lattices to stabilize this 
configuration: $12\times 24$ for $x$ = 2/9, and $N'\times 12$, with $N'$ = 48 
(1/6), 24 (1/9), and 32 (1/12) (there was no similar problem for the LAF width=2
stripes, which are also metastable). 
The surface tension also depends sensitively on the free energies of
the reference end phases.  These can be calculated either exactly, from the
mean field theory, or numerically from the UHF.  For the LAF the agreement is
quite good: energy per site $E_{LAF}/t$ = 1.46579 (mean field) vs 1.46578 (UHF);
$E_{AFM}/t$ = 2.46577 (mean field) vs 2.46588 (UHF) (UHF's on $24\times 24$
matrices).  It was necessary to use the UHF value for $E_{AFM}$ to calculate
surface tensions. 

The present calculations shed some light on the controversy in the tJ model.  
The doped AFM phase is so unstable in the Hubbard model, that it is likely that
the elementary excitations in the `uniform' lightly-doped tJ model are really
magnetic polarons.  A recent Quantum Monte Carlo study of the Hubbard 
model\cite{GEH} also finds that holes add new dispersionless bands, and do not 
uniformly dope the AFM phase.  (See also Ref.~\onlinecite{WhAff}.)  Hence, the 
three-sided debate about `uniform' 
(or magnetic polaron)\cite{HelMan} vs stripe\cite{WhiSc} vs (macroscopically) 
phase-separated\cite{pry} tJ ground state is in all probability really a debate 
about three kinds of phase separated ground state.  Our results favor 
(meandering) stripes.

In conclusion, we find WS-like stripes at the HF level in the Hubbard model
(albeit as metastable states), and we demonstrate that they arise from a 
tendency to phase separation, providing the first estimate of their surface
tension.

{\bf Acknowledgment:}  CK's research was supported in part by NSF Grant
NSF-9711910.  These computations were carried out using the facilities
of the Advanced Scientific Computation Center at Northesatern University 
(NU-ASCC).  Their support is gratefully acknowledged.  
Publication 783 of the Barnett Institute.

\end{document}